\documentstyle[eqsecnum,aps,twocolumn]{revtex}

\newtheorem{lemma}{Lemma}
\newtheorem{definition}[lemma]{Definition}
\newtheorem{theorem}[lemma]{Theorem}

\newtheorem{proposition}[lemma]{Proposition}
\newtheorem{remark}[lemma]{Remark}
\newtheorem{corollary}[lemma]{Corollary}

\newenvironment{proof}{{\bf Proof:}}{\hfill$\Box$}
\begin{document}
\draft
\preprint{HEP/123-qed}
\title{Temporary Assumptions for Quantum Multiparty Secure Computations}
\author{J. M{\"u}ller-Quade and H. Imai}
\address{Imai Laboratory, Institute of Industrial Science, The University of 
Tokyo}
\date{May $31^{st}$, $2001$}
\maketitle

%
%

\begin{abstract}
This paper introduces quantum multiparty protocols which allow the use
of temporary assumptions.  We prove that secure quantum multiparty
computations are possible if and only if classical multi party
computations work. But these strict assumptions are necessary only
during the execution of the protocol and can be loosened after
termination of the protocol.

We consider two settings:
\begin{enumerate}
\item A collusion of players tries to learn the secret inputs of
honest players or tries to modify the result of the computation.
\item A collusion of players cheats in the above way or tries to
disrupt the protocol, i.\,e., the collusion tries to abort the
computation or leaks information to honest players.
\end{enumerate}

We give bounds on the collusions tolerable after a protocol has
terminated and we state protocols reaching these bounds.
\end{abstract}

%
%

\section{Introduction}

Due to the no-go theorems of Mayers and Lo/Chau~\cite{May96,LoCha96}
and Lo~\cite{Lo96} quantum cryptography cannot---with unconditional
security---implement bit commitment, oblivious transfer and many other
important two party protocols.

Here we give an analysis of the case of multiparty protocols.

We will investigate two settings. First multiparty protocols which we
call {\em partially robust}, which can tolerate all forms of
cheating, but can be aborted by a collusion of disruptors and 
secondly we will consider multiparty protocols which are {\em robust}
even against disruption~\cite{ImaMue00Eurocrypt}.

In the case without disruption classical multiparty protocols can
yield unconditional security against all possible forms of cheating if
a majority of the players is honest and one assumes private
channels between any two parties as well as a broadcast
channel~\cite{BenGolWid88,ChaCreDam88,RabBen89}. More general for
every set of possibly colluding parties secure multiparty computations
are possible if no two collusions cover the total set of
players~\cite{HirMau97}.

From the no-go theorems for quantum two party protocols it can be
concluded that there exist functions which cannot be realized by
secure quantum multiparty protocols if two sets of possibly colluding
parties cover the complete set of players. If the possible
collusions are only defined by their cardinality multiparty protocols
using quantum cryptography become insecure if not a majority of
players behave honestly.

Still quantum multiparty protocols have advantages over classical
multiparty protocols. In this paper we prove that the assumptions
about possible collusions can be loosened after the execution of the
protocol. We present protocols where a majority may become dishonest
after the protocol has terminated. Furthermore we give a limit on the
collusions which are tolerable after the execution of the protocol.
These bounds depend on the collusions which are tolerable during the
execution of the protocol.

In the case with disruption we need that no two possible collusions
cover all but one players, i.\,e., the cardinality of the union of two
collusions never reaches $n-1$ for $|P|=n$. Given this assumption we
can again prove that the assumptions about possible collusions can be
loosened after the execution of the protocol. We also give limits on
the collusions which are tolerable after the execution of the protocol
and prove that these limits are tight.

We will restrict our view mostly to realizing a bit commitment from
one party (called Alice) to a party named Bob. For our impossibility
results we simply prove that relative to the given assumptions there
cannot exist a multiparty protocol realizing a bit commitment from
Alice to Bob. This implies that under the given assumptions there
exist functions which cannot be computed securely. For the
constructive results it is again enough to look at bit commiment,
because with a result of Yao~\cite{Yao95} we can realize an oblivious
transfer channel using a quantum channel and bit commitments. With
such an oblivious transfer channel between every pair of players we can
realize multiparty computations, even with a dishonest
majority~\cite{BeaGol89,GolLev90,CreGraTap95,ImaMue00Eurocrypt}. Our concern
will hence be to characterize the assumptions relative to which a bit
commitment between two of the players becomes possible.

The structure of the paper is as follows.  In
Section~\ref{secretSharing} we will review definitions and known
results on classical multiparty protocols and secret sharing
techniques as far as we need them to prove our results.  Next, in
Section~\ref{NoGo} we review the impossiblity of quantum bit
commitment in the two party scenario~\cite{May96,LoCha96}. We give
some generalizations to the situation of multiparty protocols. In
Section~\ref{TempAssumptions} we stress the cryptographic importance
of assumptions which can be loosened after a limited time, so called
{\em temporary assumptions}. Then in the 
Sections~\ref{ForcingMeasurements},\ref{PartiallyRobustProtocols},
\ref{RobustProtocols}
we give protocols which allow temporary assumptions in secure multiparty
computations.  The main idea is to use a classical secret sharing scheme
as a bit commitment protocol to force honest measurements. We show
that, after the honest measurements are performed, the assumptions
about possible collusions can be loosened.

%
%

\section{Classical Multiparty
  Protocols and Secret Sharing }\label{secretSharing}

\subsection{Classical Multiparty Computations}

In a multiparty protocol a set $P$ of players wants to correctly
compute a function $f(a_1,\dots,a_n)$ which depends on secret inputs
of $n$ players. Some players might collude to cheat in the protocol as
to obtain information about secret inputs of the other players or
to modify the result of the computation.

When we look at the infrastructure available for the players we are
mainly interested in three settings. First, there are private
and authenticated channels between every pair of players and each
player has a broadcast channel, second, every pair of players is
connected by an oblivious transfer channel and a broadcast channel is
available for everyone, and third, which is the setting for the
results of this paper, every two players are connected by a quantum
channel and an insecure but authenticated classical channel plus every
player has access to a broadcast channel.


In multiparty computations we have to make some
assumptions about possible collusions. We model possible
collusions by defining a set of collusions. Only one of these possible
collusions is actually cheating. Within this set of colluding
players the players share their input and take actions based on their
common knowledge.

\begin{definition}
An adversary structure is a monotone set ${\cal A}\subseteq 2^P$,
i.\,e., for a subset $S'$ of a set $S \subseteq P$ the property $S\in
{\cal A}$ implies $S' \in {\cal A}$.
\end{definition}

The main properties of a multiparty protocol are:
\begin{enumerate}
\item A multiparty protocol is said to be ${\cal A}$-{\em secure} if
no single collusion from $\cal A$ is able to obtain information about
the secret inputs of other participants which cannot be derived from
the result and the inputs of the colluding players.
\item A multiparty protocol is ${\cal A}$-{\em partially correct} if no
party
  can let the protocol terminate with a wrong result.
\item A multiparty protocol is ${\cal A}$-{\em correct} whenever no
  single collusion from $\cal A$ can abort the protocol, modify its
  result, or deviate from the protocol in a way that an honest player
  obtains information about the secret inputs of another player which
  cannot be derived from the result and the input of this honest player.
\item A multiparty protocol is called $\cal A$-{\em fair} if no
  collusion from $\cal A$ can reconstruct the result of the multi
  party computation earlier then all honest participants together. No
  collusion should be able to run off with the result.
\end{enumerate}

A multiparty protocol having the properties 1., 2. and 4. is called
${\cal A}$-{\em partially robust} and a protocol having
all three above properties is called
${\cal A}$-{\em robust}.

Whenever we are only concerned with partially robust protocols we will
abort the protocol whenever a player complains about another player.
Only robust protocols must be able to cope with conflicts between
players.

Note that we allow only one collusion from $\cal A$ to cheat.
Furthermore active cheaters are always be considered to be passively
cheating, too.





Sometimes one thinks of all players being equivalent in their
trustability, then adversary structures are solely defined by the
cardinality of the collusions.  When
refering to an adversary structure which contains all subsets of $P$
with no more than $t$ players we denote the above properties by
$n \choose t$-secure, $n \choose t$-(partially) correct,
$n \choose t$-fair, and
$n\choose t$-robust\footnote{Sometimes the terms
  $t$-secure, $t$-(partially) correct, $t$-fair, and $t$-robust are used
for
  $n\choose{t-1}$-secure, $n\choose{t-1}$-correct,
  $n\choose{t-1}$-fair, and $n\choose{t-1}$-robust}.

\subsection{Multiparty Computations with Private Channels}

We will summarize next what can be achieved by classical multiparty
computations when private channels are available between any two
players as well as a broadcast channel. The next result is
taken from~\cite{HirMau97}.

\begin{theorem}
  Given a set $P$ of players with a secure and authenticated channel
  between each pair of players together with a broadcast cannel, then
  every function can be computed by an $\cal A$-partially robust
  multiparty protocol if no two sets from $\cal A$ cover the complete
  set $P$ of players.
\end{theorem}

\begin{remark}
  There exist functions for which a multiparty protocol among players
who have access to a broadcast channel and have secure and authenticated
channels connecting every pair of players cannot be $\cal A$-robust if
two collusions cover $P\setminus \{ P_i\}$ for some player $P_i$.
\end{remark}

\begin{proof}
If the players of two possible collusions $A_1, A_2\in {\cal A}$
covering $P\setminus \{ P_i\}$ cannot cooperate then it is not clear for
$P_i$ which collusion is cheating. To continue with the protocol all
messages between players who are complaing about each other have to be
exchanged over the broadcast channel or over secure channels via
$P_i$. Obviously $P_i$ learns all secrets or the protocol must be
aborted. In both cases the protocol is not $\cal A$-robust.
\end{proof}

As a corollary we state the classic result
from~\cite{BenGolWid88,ChaCreDam88} which was generalized to the
situation with broadcast channel in~\cite{RabBen89}.

\begin{corollary}
  Given $n$ players which have a secure and authenticated channel
  between each pair of players together with a broadcast cannel, then
  every function can be computed by a $n\choose t$-partially robust
  multiparty protocol if $t<n/2$, i.\,e., if a majority is honest.
\end{corollary}

\subsection{Secret Sharing}\label{SecretSharing}

One important primitive of classical multiparty protocols is secret
sharing which was introduced in~\cite{Sha79,Bla79}. 
The aim of a secret sharing scheme is to allow a dealer to
distribute shares $s_1,s_2,\dots,s_n$, which represent one secret value
$x$, to a set $P$ of players such that only certain authorized subsets of
$P$ can reconstruct the secret
$x$ whereas all other subsets of $P$ can't get any information about $x$.
Of course every subset of $P$ containing an authorized set must also
be authorized. This leads to the definition of an {\em access structure}.

\begin{definition}
  An access structure is a set ${\cal Z}\subseteq 2^P$ for which for
  subsets $S'\supseteq S$ of $P$ the property $S\in {\cal Z}$ implies
  $S' \in {\cal Z}$.
\end{definition}

An acess structure is ``dual'' to an adversary structure. If ${\cal A}$
is an adversary structure then the set $\{ A^c | A\in{\cal A}\}$ forms
an access structure.

The first access structures which were studied were defined by the
cardinality of their minimal authorized sets.  Later secret sharing
schemes were constructed for
arbitrary access structures~\cite{ItoSaiNis87,BenLei88}.


For our protocol it is especially important to be able to keep 
the dealer from deliberately handing out faulty shares. Shares which
do not match the agreed on access structure. 

This problem can be overcome by verifiable secret
sharing~\cite{Ben86,ChaCreDam88,BetKnoOtt93,HirMau97} which allows a
verifier (e.\,g. each individual player) to check if a share he
received is a valid share.

As we will use it later we sketch a verifiable secret sharing scheme
from~\cite{Ben86}
which can be used for every homomorphic secret sharing scheme. I.\,e.,
for every scheme where the secrets form an additive group and sharing
is a group homomorphism.

\vspace*{3mm}

{\bf Verifiable Secret Sharing}$(m)$
\begin{enumerate}
\item Alice shares a secret $m$ with access structure$\cal Z$.
\item {\tt for} $j=1$ {\tt to} $k$  {\tt do}\\
\begin{enumerate}
\item Alice shares a random secret $z$ with the access structure
$\cal Z$.
\item The verifier tells Alice to either open $z$ or $z\oplus m$.
\item Alice publishes the shares for  $z$ or $z\oplus m$. 
\end{enumerate}
\item[] {\tt od}
\end{enumerate}

If no player complains about the shares Alice publishes and if the
shares published were correct shares then the verifier is convinced
that all honest players hold correct shares.

It is clear that the secret $m$ is shared correctly if $z$ is shared
correctly and $z\oplus m$ is shared correctly. A dishonest dealer will
be caught cheating with a probability of $\frac{1}{2}$ in every of the
$k$ iterations. Hence the probability to pass this test with an
incorrectly shared secret is $2^{-k}$ and thus negligible in $k$.

\subsection{Multiparty Computations with Oblivious Transfer}

Given an oblivious transfer channel all secure two party computations
become possible with unconditional security~\cite{Kil88}. This result was
generalized to allow multiparty computations with a dishonest
majority~\cite{BeaGol89,GolLev90,CreGraTap95}. One obvious problem with such
protocols is that if a majority of players cannot run off with the
secret, i.\,e., they cannot reconstruct the secret on their own, then
a minority of players can abort the protocol. For this reason we
defined a multiparty protocol to be ${\cal A}$-{\em partially correct}
if no collusion from ${\cal A}$ can make the protocol terminate with a
wrong result.

The result of~\cite{BeaGol89,GolLev90,CreGraTap95} can then be stated as

\begin{theorem}
  Given an oblivious transfer channel between any two
  players as well as a broadcast channel, then every
  function can be realized by a $\emptyset$-robust, $2^P$-secure,
  $2^P$-fair, and
  $2^P$-partially correct multiparty protocol.
\end{theorem}

In multiparty protocols the inputs are usually shared by a secret
sharing scheme and the result is computed locally on the shares and by
sharing intermediate results. In~\cite{BeaGol89,GolLev90,CreGraTap95} the
players are committed to the shares they hold. The computation
in~\cite{CreGraTap95} uses a
{\em global committed oblivious tranfer}, which is constructed there,
to implement NOT and AND gates directly on the commitments.  As the
players are unable to cheat in the global committed oblivious tranfer
and the players cannot open their commitments faultily every form of
cheating is detectable. The only problem is that it is not always clear
who is cheating. In~\cite{ImaMue00Eurocrypt} more robust protocols
based on oblivious transfer were analyzed. There is a trade off
between robustness and security as stated in the following result
which are taken from~\cite{ImaMue00Eurocrypt}.

\begin{lemma}\label{AprioriLemma}
Let $P$ be a set of $n$ players with every pair of players being
connected by an oblivious transfer channel and every
player having access to a broadcast channel. Let $\cal A$ and
$\widetilde{\cal A}$ be adversary structures, then for all functions
$\cal A$-robust and $\widetilde{\cal A}$-secure multiparty protocols exist
if
\begin{enumerate}
\item the adversary structure $\cal A$ does not contain two sets
  covering $P\setminus \{P_i\}$ for any $P_i\in P$ and
\item the adversary structure $\widetilde{\cal A}$ contains only the
complement of one previously chosen set $B$ which is maximal in $\cal A$.
\end{enumerate}
\end{lemma}

For a proof see~\cite{ImaMue00Eurocrypt}

In the above result one can see the trade off between robustness and
security. The smaller $\cal A$ can be chosen the larger $\widetilde{\cal
A}$ will be.

\begin{corollary}\label{Cor3MPwithOT}
The protocol of Lemma~\ref{AprioriLemma} for the computation of a
function $f(a_1,\dots,a_n)$ is efficient in the number of players and
the size of the circuit used to calculate $f(a_1,\dots,a_n)$.
\end{corollary}

For a proof see~\cite{ImaMue00Eurocrypt}.

%
%

\section{No-Go Results}\label{NoGo}

\subsection{Quantum Bit Commitment Is Impossible}

In this section we will shortly review the impossibility of quantum
bit commitment as proven by Mayers and Lo/Chau.

\begin{definition}
  A {\em bit commitment protocol} is a protocol consisting of two
  phases: {\em commit} and {\em unveil}. In the commit phase Bob
  obtains information from Alice which binds her to a certain bit $b$.
  In the unveil phase Alice opens $b$ to Bob and proves to Bob that
  the commitment bound her to the bit $b$.

  A bit commitment protocol must have two properties:
\begin{enumerate}
\item binding, i.\,e., after committing Alice can, without the help of
  Bob, only unveil one
  fixed bit $b$.
\item concealing, i.\,e., without the help of Alice Bob cannot know
  the bit $b$ Alice committed to.
\end{enumerate}
\end{definition}

To show the impossibility of quantum bit commitment one proceeds in two
steps
\begin{enumerate}
\item First one shows that for each quantum protocol there exists a
protocol which keeps all actions at the quantum level and postpones
all measurements and random choices until shortly before unveil. This
protocol is secure (binding and concealing) if and only if the
original protocol was secure.
\item Then it is proven that (in the new protocol) either Bob can from
  his part of the quantum state distinguish between a comitted zero
  and a committed one (the protocol is not concealing) or Alice can
  with a quantum transformation change her part of the quantum state
  from a superposition of commitments to zero into a superposition of
  commitments of one (and vice versa).
\end{enumerate}

A key insight in the impossibility proofs given
in~\cite{May96,LoCha96} was that for each quantum protocol which
involves measurements, random choices, and classical communication one can
construct an
equivalent protocol which has all measurements and random choices
postponed to shortly before the unveil phase. With this reduction it is
possible to treat the result of the commit phase as being a pure
quantum state shared by Alice and Bob.

We will shortly explain the attack in more detail. Special emphasis is
put on on classical
communication during the protocol, as it involves measurements. But
again these measurements can theoretically be delayed.

Alice behaves like she wants to honestly commit to zero and Bob
behaves like an honest Bob, but all decisions which have to be made in
the course of the protocol will no more be based on measurement
results (or random choices), but will be done by {\em conditional
quantum gates}~\cite{Gru99} hence keeping all possibilities in
superposition up to the measurement shortly before the unveil
phase. Random choices are done in the same way, instead of fixing one
value all possible values should be created in superposition.  The
most critical part of the reduction concerns classical
communication. To get this classical data one must perform a
measurement but even this measurement can be delayed to shortly before
the unveil phase without changing the security of the
protocol. Instead of measuring a qubit Alice entangles this qubit with
two new qubits such that all three qubits give the same measurement
result in the basis which should be used for the measurement (this is
done by two {\em controlled not}) and sends one of these new qubits to
Bob. This way Bob gets the information, and Alice can measure what
information Bob got, but these measurements can be postponed without
changing the security of the protocol.

If both parties follow this technique to keep everything at the
quantum level, then the protocol will deliver, after the commit phase,
a pure state shared by Alice and Bob.

This pure state $|\Psi_0\rangle$ appears on Alice side as a mixture
$\rho_{Alice,0}$ (the index zero reminds us that Alice has committed
to zero, the states $|\Psi_1\rangle$ and  $\rho_{Alice,1}$ correspond to a
commitment of one). On Bobs side the pure state appears as $\rho_{Bob}$.

Now we can import a result from~\cite{HugJozWoo93}.
\begin{theorem}[Hughston, Jozsa, and Wootters]\ \\
  Given two pure quantum states $|\psi\rangle$ and $|\phi\rangle$
  shared between Alice and Bob which appear as the same state $\rho$
  on Bob's side, then there exists a unitary transform
  $U_{|\psi\rangle,|\phi\rangle}$ which acts on Alices part of the
  quantum system only and changes $|\psi\rangle$ to $|\phi\rangle$.
\end{theorem}

This result was generalized by Mayers to the case where $|\psi\rangle$
and $|\phi\rangle$ do not appear as the same state on Bob's side but
as states which are very close to each other~\cite{May95}.

In the case of bit commitment this says that either the bit can be
measured on Bob's side, i.\,e., $\rho_{Bob}$ looks different for
$|\Psi_0\rangle$ and $|\Psi_1\rangle$, or Alice can change from
$|\Psi_0\rangle$ to $|\Psi_1\rangle$ by a unitary transform ${\cal
  U}_{0,1}$ on her part of the quantum system.

We can conclude the impossibility result from~\cite{May96,LoCha96}.
\begin{theorem}[Mayers, Lo/Chau]
A quantum protocol for bit commitment cannot be binding and concealing.
\end{theorem}

 To cheat in the actual protocol it is of
course not necessary that both parties keep their decisions at the
quantum levels. It is enough if the party being able to cheat does
so. See also Lemma~\ref{whatmaybemeasured}.

\subsection{Bounds on Tolerable Adversary Structures During the Execution of
  a Protocol}

With the impossibility results for the two party case one can as well
show that quantum cryptography cannot enhance classical bounds for the
set of tolerable adversaries~\cite{HirMau97}.

\begin{corollary}\label{honestMaj}
  Let $P$ be a set of players and let ${\cal A}$ be an adversary
  structure. If there exist two possible collusions $A_1,A_2\in{\cal
  A}$ with $A_1\cup A_2 = P$ then not all functions can be computed
  ${\cal A}$-partially robustly by a quantum multiparty protocol.
\end{corollary}

\begin{proof}
  We show that it is impossible to realize a bit commitment for a
  party Alice $\in A_1$ and Bob $\in A_2$. This is simple as we are
  almost in the two party scenario: Assume the collusion $A_2$ can by
  no means measure the bit Alice committed to, then the collusion
  $A_1$ can, by keeping every action at the quantum level cheat
  analogously to the two party situation, i.e., there exists a unitary
  transform $U_{0\rightarrow 1}$ which can change the bit Alice is
  committed to. The transform $U_{0\rightarrow 1}$ must be jointly
  applied by all players in $A_1$
\end{proof}

\begin{corollary}
There exist functions which cannot be computed $n\choose t$-partially robustly
by a quantum multiparty protocol if $t\geq n/2$.
\end{corollary}

If we consider robustness we have to take into account more deviations
from the protocol. A collusion of players could for example leak their
secret (quantum) data to a player not in the collusion. Such an attack
further limits the set $\cal A$ of possible collusions.

\begin{corollary}\label{Allbutone}
  Let $P$ be a set of players and let ${\cal A}$ be an adversary
  structure. If there exist two possible collusions $A_1,A_2\in{\cal
  A}$ with $A_1\cup A_2 = P\setminus \{P_i\}$ for any player $P_i$,
  then not all functions can be computed ${\cal A}$-robustly by a
  quantum multiparty protocol.
\end{corollary}

\begin{proof}
Assume there exists a $P_i\in P$ with $P = A_1\cup A_2\cup \{ P_i\}$ for
$A_1,A_2\in{\cal A}$. We would like to implement a bit commitment from a
player from $A_2$ to the player $P_i$. To prevent the players from $A_2$
to jointly change the committed bit it must be possible for the
players of $A_1\cup\{ P_i\}$ to measure the committed bit. Only the
assumption that $P_i$ does not collude with the players of $A_1$ makes
this attack impossible. If the player $P_i$ is honest but curious and
keeps everything at the quantum level, then he would be able to
measure the committed bit if all players of the set $A_1$ would together
keep all their actions at the quantum level and later on give all
their quantum information to the player $P_i$.

Even though the player $P_i$ does not collude with the players from
$A_1$ we cannot keep the players from $A_1$ from deviating from the
protocol in giving away their secret data. 
\end{proof}

\subsection{Bounds on Tolerable Adversary Structures After the
  Protocol Terminated}

During the execution of a protocol we must use the same assumptions as
in classical multiparty protocols to obtain unconditional 
security. We will next prove bounds on the set of tolerable collusions
after a protocol has been finished. In our case: after the commit
phase of a commitment protocol has terminated. Interestingly these
bounds are different.

To apply the attack of Mayers and Lo/Chau Alice need not keep every
action at the quantum level. She can perform measurements which
yield not enough information to, together with the quantum
information Bob has, be able to distinguish between the commitments
zero and one. In short Alice can perform any measurement whose result
she could tell Bob without giving away her secret commitment.

\begin{lemma}\label{whatmaybemeasured}:
Let $|\Psi_b \rangle$ be a pure quantum state shared between Alice and
Bob which is the result of a quantum bit commitment protocol which was
executed at the quantum level. 

If Alice can change the bit she
committed to by a unitary transform $U_{0\rightarrow 1}$ on her part
of $|\Psi_b \rangle$  then she can
still change the bit after she performed a measurement on her part of
the quantum state if the information obtained by this measurement
together with the quantum information Bob holds does not allow to
distinguish between the commitments zero and one.
\end{lemma}

\begin{proof}
One can define a bit commitment protocol where Alice has to
perform this measurement and send the information measured to Bob.
As Bob can still not distinguish between the commitments zero and
one the attack of Mayers, Lo/Chau applies and
there exists a unitary transform changing the bit.
\end{proof}

This simple result helps us to prove that temporarily having an honest
but curious third party does not allow us to implement bit
commitment~\cite{ImaMue00SITA}.

\begin{lemma}\label{ThirdParty}
  Bit commitment may be implemented between Alice and Bob if we
  introduce a trusted third party, but the assumption of having an
  honest but curious third party is not a temporary assumption.
\end{lemma}

\begin{proof}
If the honest but curious party remains independent of the two parties
Alice and Bob bit commitment can be implemented by classical
multiparty protocols.

Now assume the honest but curious third party joins Alice or Bob after
the commit phase is completed.

The honest but curious third party will follow the protocol, but leave
everything in superposition which need not be sent away as classical
data. So the third party will perform some measurements. The third
party can join Bob afterwards and Bob should still be unable to
recover Alices bit.  Hence the third party did only obtain measurement
results which are of no use for Bob. Hence if the third party joins
Alice we are in the situation of Lemma~\ref{whatmaybemeasured}. Alice
together with the third party can jointly perform a unitary transform
which changes the bit Alice committed to.
\end{proof}


Lemma~\ref{ThirdParty} can be generalized to the multiparty scenario.

\begin{proposition}\label{MPNoGo}
  There exist functions for which no quantum multiparty protocol,
  which is partially robust against the adversary structure ${\cal A}$
  can afterwards become secure against an adversary structure
  which contains two complements of sets in ${\cal A}$.
\end{proposition}

\begin{proof}
  Let $A_1, A_2\in {\cal A}$ denote two sets of possibly colluding
  players and let $\widetilde{\cal A}$ be an adversary structure
  containing the complements of $A_1$ and $A_2$.  We show that it is
  impossible to implement an oblivious transfer from Alice $\in A_1$
  to Bob $\in A_2$
  which is $\cal A$-partially robust and $\widetilde{\cal A}$-secure after
  termination. Assume such an oblivious transfer were possible then we
  could with it implement a bit commitment from Alice to Bob which is
  $\cal A$-partially robust during the commit phase and
  $\widetilde{\cal A}$-partially robust up to the unveil phase. This
  is easy to see as after termination of the commit phase security is
  the only critical issue. The data computed during the commit phase
  cannot be changed any more and fairness is not of interest until the
  unveil phase.

  It remains to be proven that a bit
  commitment from the player Alice $\in A_1$ to the player Bob $\in
  A_2\setminus A_1$ is impossible.  We look at the sets $A_1$,
  $A_2\setminus A_1$ and $P\setminus (A_1\cup A_2)$ and prove the
  impossibility analogously to Lemma~\ref{ThirdParty}.  During the
  execution of the commit phase the protocol is ${\cal A}$-partially robust
  hence we can assume that the players in $A_1$ or the players in
  $A_2\setminus A_1$ collude and we still get a valid commitment from
  $A_1$ to $A_2\setminus A_1$. Now we assume the protocol to become
  $\widetilde{\cal A}$-partially robust afterwards. Then all players from
  $P\setminus (A_1\cup A_2)$ may join the players from $A_2\setminus
  A_1$ and the bit commitment remains concealing even if the players
  from $A_2\setminus A_1$ were colluding. Hence all quantum
  information in the posession of the players from $P\setminus
  (A_1\cup A_2)$ are of no use to Bob. According to
  Lemma~\ref{whatmaybemeasured} the players from $A_1\cup P\setminus
  (A_1\cup A_2) = P\setminus A_2 = A_2^c$ can change the committed
  bit. Hence the bit commitment is not $\widetilde{\cal
    A}$-partially robust between commit and unveil.
\end{proof}

\begin{corollary}
  There exist functions for which no ${n}\choose{t}$-robust quantum
  multiparty protocol can become ${n}\choose{n-t}$-robust after its
  execution.
\end{corollary}

%
%

\section{Temporary Assumptions}\label{TempAssumptions}

Usually assumptions have to be made very carefully, because they
implicitely try to predict future developements. The assumptions must
be valid as long as the secret information is critical.

Temporary assumptions are hence very promising.
There was little research into
temporary assumptions in quantum cryptography
after it became clear that computational assumptions cannot be used only
temptorarily~\cite{BraCreMaySal98}.

But quantum cryptography allows assumptions which are independent of
computational assumptions. Such assumptions can be temporary.

The key idea to get temporary assumptions is to not try to make the
transformation ${\cal U}_{0,1}$, which can change the committed bit,
impossible, but to make it impossible for the parties (at least for the
party able to cheat) to keep all actions at the quantum level.

E.\,g. Alice can trivially not cheat in the protocol of~\cite{BenBra84}
if she has no quantum storage, even if quantum storage became
available to her after the commit phase.

Assumptions which have the same effect are: limited quantum storage
capacity and limited storage time for quantum bits as well as
assumptions about decay introducing errors.  Such assumptions need
only hold during the execution of the protocol.

%
%

\section{Forcing Measurements with Secret Sharing}\label{ForcingMeasurements}

In~\cite{Yao95} Yao proved that it is possible to obtain oblivious
transfer from a {\em black box} bit commitment and a quantum
channel. The idea goes back to Crepeau~\cite{Cre94} and was
generalized to quantum channels which can have noise by
Mayers~\cite{May96b}. 

The basic idea is to force measurements to avoid the attacks of Mayers
and Lo/Chau and Lo~\cite{May96,LoCha96,Lo96}. In the course of the
protocol one party has to commit to the measurment bases used and to
the results obtained. Then a random subset of these measurments are
opened. If there are not too many discrepancies one can be sure that
the committing party did measure most of the qubits. This already
suffices to make the delay of all measurements impossible hence
avoiding the attacks of Mayers and Lo/Chau and
Lo~\cite{Yao95,May96b}.  But one has to be careful if the bit
commitment used is strong enough to force measurements. The {\em
unconditionally secure bit commitment} of Kent~\cite{Ken98} is not
suitable as Kent proved in~\cite{Ken99CCBC}.

The main requirement for a bit commitment to be able to force
measurments is that committing to a bit $b$ must be equivalent to 
giving the classical bit $b$ to a trusted third party. Committing to a
measurement result according to~\cite{Yao95,May96b} implies an
irreversible measurement as otherwise the cheater and the trusted
third party together could violate the Heisenberg uncertainty.

In this section we will show that secret sharing can be used like a
{\em black box} bit commitment to force measurements using the protocols
of~\cite{Yao95,May96b}. 

If we use, instead of bit commitment, secret sharing
with an access structure $\cal Z$ and let
$\cal A$ be the set $\{ A | A^c\in{\cal Z}\}$
then we have the following properties:
\begin{enumerate}
\item the bit commitment based on secret sharing is concealing given
only a collusion of $\{A | A\not\in {\cal Z}\}$ is cheating.
\item The bit commitment based on secret sharing is binding if only
one collusion of $\cal A$ is cheating.
\item The bit commitment based on secret sharing is equivalent to
announcing the bit to a trusted third party whenever only one
collusion of $\cal A$ is cheating.
\end{enumerate}

The first two points of this enumeration follow directly from the
properties of secret sharing schemes. Now we look at the third
point. According to the assumption that only one collusion of $\cal A$
cheats we know that there exists a set $M$ of honest players able to
reconstruct the shared secret. As all players of $M$ are honest the
committing player (Alice) had to honestly transmit all the shares of
the players of $M$. These shares already fix the committed bit and
hence handing out those shares is equivalent to announcing the bit to
a trusted third party. From this the next result follows without
further proof.

\begin{lemma}\label{ForcingviaSS}
Let $\cal A$ be an adversary structure and let $\cal Z$ be an access
structure such that ${\cal A} = \{ A | A^c\in{\cal Z}\}$. Then secret
sharing with access structure $\cal Z$ can be used to obtain $\cal
A$-partially robust oblivious transfer from any player to the dealer.
The protocol is $\{A | A\not\in {\cal Z}\}$-secure.
\end{lemma}

After the measurements are irreversibly performed and the quantum
attacks are impossible the bit commitment used
need not be binding any more. Only the concealing property is still
needed. For secret sharing the requirements for binding and
concealing are different as seen in the enumeration above.
So after all measurements are performed, especially after termination
of the protocol, only collusions from $\{A | A\not\in {\cal Z}\}$ can
cheat in the oblivious transfer.

Of course we want oblivious transfer not only from one party to a set
of players, but between every pair of players. The next section will
give a detailed analysis of this situation.

\section{Partially Robust Protocols for Oblivious Transfer}\label{PartiallyRobustProtocols}

We will next give a detailed analysis of the situation where we have a
set $P$ of players together with an adversary structure $\cal A$ and
every player should be able to share a secret among the other players.

We are only concerned with partially robust protocols here. Whenever a
player complains about another player we will abort the
protocol. Robust protocols will be presented in the next section.

\begin{lemma}\label{BCviaSS}
Let $P$ be a set of players for which each pair of players is
connected by an authenticated secure channel and every player has
access to a broadcast channel. Let $\cal A$ be an adversary structure
for which no two collusions cover the set $P$ of
players. Then a bit commitment between any pair of players is possible
which is $\cal A$-partially robust and $\{A^c | A\not\in {\cal A}
\}$-secure.
\end{lemma}

\begin{proof}
We will let Alice commit to a bit string $m\in\{0,1\}^k$.

\vspace*{3mm}

{\bf Commit via Secret Sharing}($m$)
\begin{enumerate}
\item Alice sends Bob a random string $r\in\{0,1\}^k$.
\item Alice shares the string $m\oplus r$ using a secret sharing scheme
with access structure ${\cal Z}= \{ Z | Z^c\not\in{\cal A}\}$.
\end{enumerate}

This protocol shares Alices secret $m$ with the access structure
${\cal Z}\cap \{M\subseteq P | {\rm Bob}\in M\}$.  If the receiver Bob
is honest this protocol can be used to force measurements $\cal
A$-partially robustly (Lemma~\ref{ForcingviaSS}) and if Bob is not honest
then we cannot prevent a dishonest sender from colluding and  changing the
committed bit together with the receiver of the bit commitment. No bit
commitment scheme can.

The unveil protocol is essentially a reconstruction of the shared
secret. 

\vspace*{3mm}

{\bf Unveil}
\begin{enumerate}
\item Alice announces the shares she sent.
The players from $P$ confirm the shares and Bob can then reconstruct
$m$ from his knowledge of $r$. 
\end{enumerate}
\end{proof}

We can improve the security a little bit further by not allowing every
player to commit a bit via secret sharing.  To obtain oblivious
transfer between every pair of players it is enough that for every
pair of players one of them can commit to the other as oblivious
transfer can be inverted~\cite{CreSan91}.

\begin{lemma}\label{BCviaSS2}
Let $P$ be a set of players for which each pair of players is
connected by an authenticated secure channel and every player has
access to a broadcast channel. Let $\cal A$ be an adversary structure
for which no two collusions cover the set $P$ of players and
let $M$ be any maximal set in $\cal A$. Then for every pair of players
a bit commitment is possible for one of the players to the other
player which is $\cal A$-partially robust and
$\{A^c | A\not\in {\cal A} \}\cup M^c$-secure.
\end{lemma}

\begin{proof}
The partial robustness is the same as claimed by Lemma~\ref{BCviaSS}
so we need to prove only the improved security.

We have to see first that $\{A^c | A\not\in {\cal A} \}\cup M^c$
is an adversary structure. The set $\{A^c | A\not\in {\cal A} \}$ is
an adversary structure and it contains all proper subsets of
$M^c$. Hence the set $\{A^c | A\not\in {\cal A} \}\cup M^c$ is an
adversary structure, too.

To obtain the higher security we choose for every pair of players one
player who shall commit to the other.
Let Alice and Bob be a pair of players for which either Alice, Bob$\in
M$, Alice, Bob$\not\in M$ or Alice$\in M$ and Bob$\not\in M$. Otherwise
exchange the names of the players.

We will see that the bit commitment from Lemma~\ref{BCviaSS} between
any two players Alice and Bob is $\{A^c | A\not\in {\cal A} \}\cup
M^c$-secure if used in the above defined direction. As
Lemma~\ref{BCviaSS} already proves the $\{A^c | A\not\in {\cal A}
\}$-security we are left with proving the security against the
possible collusion $M^c$.

A collusion can only cheat if it is an authorized set able to recover
a shared secret and if it contains the receiver of the bit
commitment as the shared secret is encrypted by a key $r$ only known
to the sender and the receiver of the bit commitment. If the collusion
contains the sender of the bit commitment then the committed bit can
already be derived from the inputs of the colluding players and no
security is lost. 

The direction of the bit commitment is chosen in a way that $M^c$
either contains the sender of the bit commitment or it does not
contain the receiver of the bit commitment and hence $M^c$ is not
able to reconstruct a secret bit in the bit commitment protocol.
\end{proof}

From Lemma~\ref{ForcingviaSS}, Lemma~\ref{BCviaSS2} ,
and~\cite{CreSan91} we get the following result about oblivious
transfer.

\begin{corollary}\label{OTviaSS}
Let $P$ be a set of players for which each pair of players is
connected by an authenticated secure channel and every player has
access to a broadcast channel. Let $\cal A$ be an adversary structure
for which no two collusions cover the set $P$ of players and let $M$
be any maximal set in $\cal A$. Then an oblivious transfer is possible
between every pair of players which is $\cal A$-partially robust and
$\{A^c | A\not\in {\cal A} \}\cup M^c$-secure.
\end{corollary}

\section{Robust Protocols for Oblivious Transfer}\label{RobustProtocols}

In this section we will additionally consider players who try to
disrupt the bit commitment protocol. In addition to the forms of
cheating partially robust protocols can cope with we have that
some players can leak out information to players not contained in
their collusion or some players can claim that some other players
do not follow the protocol or they can themselves refuse to send or to
receive messages. 

If one looks at a secret sharing scheme step by step the only
deviations of the protocol possible, which do not immediately give
away the identity of the disruptor, are:

\begin{enumerate}
\item Some players might not keep their shares secret.
\item Some players complain that the sender presents different shares
in the reconstruction phase then these players originally received.
\item Some players can claim to not receive any proper shares,
e.\,g. empty shares.
\end{enumerate}

This enumeration remains complete even if we consider verifiable
secret sharing as we essentially iterate secret sharing together with
some local computations (see Subsection~\ref{SecretSharing}).

If a disruption yields that the sender and the receiver of a bit
commitment are in conflict then we will abort this bit commitment.
We will not yet realize bit commitments between players who are
in conflict with each other. We will later use multiparty protocols to
obtain bit commitment and oblivious transfer between players who are
in conflict.

Next we will give a bit commitment scheme based on verifiable secret
sharing which can cope with disruption. Shares of complaining players
will be published, but we will see that this does not harm the security.
Thye number $k$ is a security parameter.

\vspace*{3mm}

{\bf Commit via Secret Sharing}$(m)$
\begin{enumerate}
\item Alice shares the secret $m\oplus r$ with the access structure
$\cal Z$ and sends $r$ to Bob.
\item A set $A$ complains about the shares they receive. These shares
will be published by Alice
\item $A_{old} := A$
\item {\tt repeat}
\begin{enumerate}
\item{\tt for} $j=1$ {\tt to} $k$  {\tt do}\\
\begin{enumerate}
\item Alice shares a random secret $z$ with the access structure
$\cal Z$.
\item Bob tells Alice to either open $z$ or $z\oplus m\oplus r$.
\item Alice publishes the shares for  $z$ or $z\oplus m\oplus r$. A
set $A_j$ of players complains about these shares.
\item $A := A \cup A_j$
\end{enumerate}
\item[] {\tt od}
\item If $A\not\in{\cal A}$ then Alice is detected cheating else Alice
has to publish the shares for the secret  $m\oplus r$ for all players
in $A$.
\end{enumerate}
\item {\tt until} $A_{old} = A$.
\end{enumerate}

\begin{lemma}\label{PublishShares}
Let $P$ be a set of players and let $\cal A$ be an adversary structure
for which no two collusions cover the set $P\setminus \{P_i\}$ for any
player $P_i$ and let $M$ be any maximal set in $\cal A$. Then a secret
which is shared among the players of $P$ according to the above
protocol with access
structure ${\cal Z}= \{ Z | Z^c\in {\cal A}\}$ remains to be $\{A^c |
A\not\in {\cal A} \}\cup M^c$-secure even if a collusion of $\cal A$
publishes their shares.
\end{lemma}

\begin{proof}
The security of a bit commitment scheme is only relevant if the sender
is honest. So we can assume throughout the proof that the secret is
properly shared. 

Because no two collusions of $\cal A$ cover $P\setminus\{ P_i\}$ every
set of the access structure ${\cal Z} = \{ Z | Z^c \in{\cal A}\}$
contains at least two honest players. So even if all players of a
collusion $A\in{\cal A}$ leak their shares the secret remains shared
among at least two honest players and no single honest but curious
player gets to know a secret.
\end{proof}

\begin{lemma}\label{RobustBCviaSS}
Let $P$ be a set of players for which each pair of players is
connected by an authenticated secure channel and every player has
access to a broadcast channel. Let $\cal A$ be an adversary structure
for which no two collusions cover the set $P\setminus \{P_i\}$
for any player $P_i$ and
let $M$ be any maximal set in $\cal A$. Then for every pair of players,
who will not be in conflict after the protocol, a bit commitment is
possible from one of the players to 
the other player which is $\cal A$-robust and
$\{A^c | A\not\in {\cal A} \}\cup M^c$-secure.
\end{lemma}

\begin{proof}
We are proving our claim using the above protocol.  Whenever the
receiver of the bit commitment complains about the sender then we need
not be able to implement a bit commitment hence in the following we
analyse only conflicts between the sender of the bit commitment and
players other than the receiver.

We consider two cases.  

First: Alice is honest, then
every complaint about Alice comes from a cheater and Alice publishes
the share the cheater complained about.  So every complain about an
honest Alice is equivalent to a leak of the share of the cheating
player and we have seen that this does not harm the security of the
secret sharing (Lemma~\ref{PublishShares}). The correctness of the
secret sharing, which implies the binding property of the bit
commitment, cannot be harmed by shares which become publicly known.

Second: Alice is dishonest. Then Bob must be honest or we cannot
expect a bit commitment to work. All shares the public ones as well as
the only privately known ones pass the verifiable secret sharing test
whenever the protocol has terminated and Alice has not been detected
cheating. 
Hence every honest player is convinced that if Bob is honest then the
secret is properly shared. This follows directly from the properties
of the verifiable secret sharing scheme of~\cite{Ben86} which was
sketched in Subsection~\ref{SecretSharing}. The security of the protocol
is not an issue if Alice is dishonest.
\end{proof}

From the same arguments as used in Lemma~\ref{ForcingviaSS} it is
clear that the above bit commitment can be used to force
measurements. As the direction of an oblivious transfer can be
inverted~\cite{CreSan91} the directions of the bit commitments do not
matter any more. Hence we can conclude the following.

\begin{corollary}\label{OTviaQC}
Let $P$ be a set of players for which each pair of players is
connected by a quantum channel and an authenticated insecure channel
and every player has access to a broadcast channel. Let $\cal A$ be an
adversary structure for which no two collusions cover the set
$P\setminus \{P_i\}$ for any player $P_i$ and let $M$ be any maximal
set in $\cal A$. Then for every pair of players, who will not be in
conflict after the protocol, an oblivious transfer is possible
which is $\cal A$-robust and $\{A^c
| A\not\in {\cal A} \}\cup M^c$-secure.
\end{corollary}

In the next section we will use the oblivious transfer of this
corollary to implement multiparty computations.

\subsection{Main Results}

This section is separated in two subsections the first considering
only partially robust protocols, i.\,e., protocols which are aborted
whenever a conflict occurs and the second subsection deals with robust
protocols tolerating every form of cheating and disruption.

\subsection{Partially Robust Protocols}

Combining the results of Corollary~\ref{honestMaj},
Corollary~\ref{OTviaSS}, and~\cite{CreGraTap95} we get the following
result without further proof.

\begin{theorem}\label{MainResult1}
  Let $P$ be a set of players each having access to a broadcast
  channel and let every pair of players of $P$ be connected by a
  quantum channel and an insecure but authenticated classical
  channel. 
  Then $\cal A$-partially robust  quantum multiparty protocols 
  for all functions exist if and only if no two collusions of $\cal A$
  cover $P$.

  These protocols are $\widetilde {\cal A}$-secure after termination
  if and only if the adversary structure $\widetilde{\cal A}$ contains
  at most one complement of a previously chosen set from ${\cal A}$.
\end{theorem}

Secret sharing need not be efficient, but all other protocols used
require only polynomial resources.

\begin{corollary}
  If secret sharing can be implemented efficiently for the access
  structure ${\cal Z}=\{A^c | A\in {\cal A}\}$ then the protocols of
  Theorem~\ref{MainResult1} can be efficient.
\end{corollary}

\subsection{Robust Protocols}

To obtain robust protocols we must according to Lemma~\ref{Allbutone}
choose $\cal A$ such that no two possible collusions together
contain all but one players. In this situation we can use
Lemma~\ref{AprioriLemma} together with Corollary~\ref{OTviaQC} and obtain the
following result without further proof.

\begin{theorem}\label{MainResult2}
  Let $P$ be a set of players each having access to a broadcast
  channel and every pair of players of $P$ being connected by a
  quantum channel and an insecure but authenticated classical
  channel. 
  Then $\cal A$-robust  quantum multiparty protocols 
  for all functions exist if and only if no two collusions of $\cal A$
  cover $P\setminus\{ P_i\}$ for any player $P_i$.

  These protocols are $\widetilde {\cal A}$-secure after termination
  if and only if the adversary structure $\widetilde{\cal A}$ contains
  at most one complement of a previously chosen set from ${\cal A}$.
\end{theorem}

If we are interested in robustness after termination and not security
after termination we can use the following result
from~\cite{ImaMue00Eurocrypt} (for a proof
see~\cite{ImaMue00Eurocrypt}).

\begin{lemma}
A multi party protocol which is $\cal A$-secure after termination is
$\cal B$-robust after termination for ${\cal B} = \{ B | \exists
A\in{\cal A}:B\subset A {\rm and} B\not= A \}$.
\end{lemma}

Again we need only polynomial resources if secret sharing is
efficient.

\begin{corollary}
  If secret sharing can be implemented efficiently for the access
  structure ${\cal Z}=\{A^c | A\in {\cal A}\}$ then the protocols of
  Theorem~\ref{MainResult2} can be efficient.
\end{corollary}

\end{document}